# Mechanical Performance of 3D Printed Interpenetrating Phase Composites with Spinodal Topologies

Yunfei Zhang[a], Meng-Ting Hsieh[b], Lorenzo Valdevit [a,b*]

[a] Department of Materials Science and Engineering, University of California, Irvine, CA 92697, USA

[b] Department of Mechanical and Aerospace Engineering, University of California, Irvine, CA 92697, USA

**Abstract**

The mechanical response of interpenetrating phase composites (IPCs) with stochastic spinodal topologies is investigated experimentally and numerically. Model polymeric systems are fabricated by Polyjet multi-material printing, with the reinforcing phase taking the topology of a spinodal shell, and the remaining volume filled by a softer matrix. We show that spinodal shell IPCs have comparable compressive strength and stiffness to IPCs with two well-established periodic reinforcements, the Schwarz P triply periodic minimal surface (TPMS) and the octet truss-lattice, while exhibiting far less catastrophic failure and greater damage resistance, particularly at high volume fraction of reinforcing phase. The combination of high stiffness and strength and a long flat plateau after yielding makes spinodal shell IPCs a promising candidate for energy absorption and impact protection applications, where the lack of material softening upon large compressive strains can prevent sudden collapse. Importantly, in contrast with all IPCs with periodic reinforcements, spinodal shell IPCs are amenable to scalable manufacturing via self-assembly techniques.

# 1. Introduction

Composite materials attain their superior combinations of properties from the synergistic contribution of each constituent phase. In the vast majority of structural composites, only one phase (the reinforcement) is stiff and strong and responsible for load bearing, while the other phase (the matrix) provides load transfer between elements of the reinforcement phase and guarantees structural stability of the composite. In these traditional designs, only the matrix phase is continuous throughout the component, whereas the reinforcement elements (fibers, whiskers, particulates…) are topologically isolated. While these designs can be very efficient under specific loading conditions (e.g., unidirectional fiber-reinforced laminae under tensile uniaxial load), their efficiency significantly suffers under more complex multi-axial states, where the lack of a continuous reinforcement phase inevitably introduces soft and weak directions.

∗ Corresponding author. Tel.: +1 949 824 4938
E-mail address: valdevit@uci.edu (L.Valdevit).

Interpenetrating phase composites (IPCs), composed of two phases that are topologically interconnected throughout the microstructure [1], represent an exciting alternative to traditional designs and have drawn a lot of research interest in recent years. Multiple studies, both numerical [2]–[10] and experimental [11]–[18], have consistently demonstrated that the combination of reinforcement and matrix in such a way that both phases are independently self-supporting and load-bearing results in improved mechanical properties compared to traditional discontinuously reinforced composite materials as the topological interconnectivity allows each constituent phase to contribute its most desirable properties to the overall properties of the composite. The synergistic role of two phases can be engineered to result in intriguing combinations of properties. As a few examples, (i) in $Al_2O_3$/Al IPCs, the stiffness of the composite was shown to be superior than that of Al, while its toughness and structural integrity were superior than that of monolithic $Al_2O_3$ [19]; (ii) while traditional aluminum foams undergo irrecoverable plastic deformation throughout their stress plateau, aluminum foam–polyurethane IPCs display extensive recoverable deformation under cyclic loadings, thanks to the stabilizing effect of the soft elastomeric phase [17]; (iii) metal-polymer interpenetrating phase nanocomposites based on Ti nanoporous can be tuned to match the elastic modulus of human bones, showing great potential for implant applications [18]; (iv) finally, molecular dynamics simulations have indicated that silicon carbide–aluminum IPCs display unique toughening mechanisms [10]. All these studies have consistently shown that IPCs are promising candidates for structural and multi-functional applications. It is worth noting that the vast majority of IPCs in these studies are manufactured through conventional methods, such as powder metallurgy [12] and infiltration processes [13], [20]. While efficient and generally scalable, these approaches randomly distribute controlled amounts of the two phases, and do not allow full topological control of the composite architecture.

Recent advancements in additive manufacturing have made it possible to manufacture IPCs with controlled and complex topologies [21]–[25]. Notable examples that have been recently extensively investigated are IPCs based on triply periodic minimal surfaces (TPMS). TPMS are mathematically defined periodic surfaces which have zero mean curvature everywhere on the surface, resulting in locally minimal surface area in each defined unit cell. Examples of TPMS include the Schwarz P surface, the Schwarz D surface and the gyroid surface [26]. In recent years, the mechanical performance of TPMS-based structures has drawn much research interest. For example, cellular materials based on TPMS have been shown to be mechanically efficient compared to truss-based cellular materials, by virtue of their smooth and regular topology, resulting in low local stress concentrations and hence efficient load transfer [26] ,[28]. Recent studies have shown that similar benefits in mechanical properties extend to TPMS-based IPCs, where either the interface between the two solid phases is a TPMS or the reinforcement phase is a thickened TPMS shell structure, embedded in a softer matrix. In particular, excellent combinations of high stiffness, strength and energy absorption have been demonstrated [29]–[34], and TPMS-based IPCs have been shown to possess better mechanical properties than their truss-based counterparts [31]. In addition, TPMS-based IPCs have been shown to possess multifunctional extrema [35], [36]. However, due to the periodic nature of TPMS, all TPMS-based IPCs are generally difficult to manufacture in a scalable fashion, thus limiting their potential applications. Far better scalability can be achieved by incorporating a stochastic reinforcement phase that can be self-assembled through spinodal decomposition of two materials followed by material conversion.

Spinodal decomposition is a thermodynamic transformation where a homogeneous (solid or liquid) solution separates spontaneously into two coexisting phases [37]. The result is an interpenetrating phase composite with very large interfacial area between the phases, and hence a characteristic length scale (domain size) much smaller than the sample size; while thermodynamics tends to reduce the interfacial area between the two phases (hence increasing the characteristic length scale), this growth can be arrested by reducing the temperature of the system (in the case of solid phases) [38] or by jamming the interface with particles that are immiscible in both phases (in the case of liquid phases) [39], [40]. A number of materials conversion techniques can be subsequently used to (i) eliminate one of the phases and converting the remaining phase to the desired material (hence producing a cellular material with spinodal solid topology) [41], (ii) converting both phases to the desired materials (resulting in an IPC with spinodal solid topology) [42], [43], or (iii) eliminating one phase, coating the other phase with the desired material and finally eliminate the second phase as well (resulting in a cellular material with spinodal shell topology) [40] . As a notable example, in a recent study bicontinuous interfacially jammed emulsion gels (bijels) are formed and processed into sacrificial porous nickel scaffolds for chemical vapor deposition to produce freestanding three-dimensional turbostratic graphene (bi-3DG) monoliths with spinodal shell topologies, possessing exceptionally high specific surface area and exceeding 100,000 unit cells [44]. In all cases, the inherent self-assembly of spinodal topologies provides a route to fabricate micro- or nano-architected materials with macroscopic dimensions, with a level of scalability unmatched by any additive manufacturing technique [44], [45]. As spinodal topologies are bicontinuous, such micro/nano-architected materials could be infiltrated with a second phase via deposition or infiltration processes, potentially providing a uniquely scalable fabrication process for shell-based IPCs.

Previous studies have shown that cellular materials with spinodal shell topologies are exceptionally efficient from the perspective of specific stiffness and strength, when compared to both cellular materials with spinodal solid topologies and truss-based lattice materials [46], and perform on par with cellular materials with triply periodic minimal surface (TPMS) topologies [47], [48]. The similarity with TPMS topologies can be attributed to the very tight distribution of curvatures in spinodal topologies, with the vast majority of surface patches possessing near-zero mean curvature and negative Gaussian curvature, the signature feature of TPMSs. Remarkably, the intrinsically stochastic nature of spinodal shell topologies (and hence the deviation from minimal surface characteristics) not only does not significantly depress mechanical properties, but rather results in considerable imperfection insensitivity [46]. Whether the unique combination of scalable manufacturing and mechanical performance shown by cellular materials with spinodal topologies also translates to interpenetrating phase composites with spinodal topologies remains to be demonstrated.

In this paper, we investigate the mechanical properties of interpenetrating phase composites with spinodal shell reinforcement, with emphasis on stiffness, yield strength and energy absorption under non-linear deformations. Spinodal shell-based IPCs are compared with composites with other reinforcement topologies, namely (1) spinodal solids, where both phase topologies are directly obtained by spinodal decomposition [42], [43]; (2) octet lattices, the most widely studied cellular architecture [49]–[51]; and (3) Schwarz P surfaces, one of the most mechanically efficient TPMS topologies [30], [52]. While a wide range of truss lattice and TPMS topologies exist, and many have been characterized mechanically, the octet lattice and the Schwartz P surface approach upper bounds of performance for the two classes of periodic reinforcement topologies, respectively, and are thus ideal candidates for assessing the mechanical efficiency of spinodal shell reinforcements. In order to accurately control all topologies, all IPCs in this study are produced by multi-material jetting, a recently developed additive manufacturing technique. While the resulting materials are polymer/polymer composites with stiffness and strength far below those of any structural material, the significant difference in mechanical properties between the two constituents allows extraction of mechanistic understanding that can be readily extend to other classes of composite materials, including ceramic/metal composites.

## 2. Materials and methods

### 2.1 Numerical generation of the reinforcement topologies

#### 2.1.1 Spinodal solid topology

The spinodal solid topology was generated numerically with the approach reported in [46] and detailed in Appendix A. A short synopsis of the procedure is presented here. A 50% dense spinodal solid cellular topology is generated by solving the Cahn–Hilliard evolution equation [37], one of the classic evolution models for spinodal decomposition. The Cahn–Hilliard equation can be written as:

$$\frac{\partial u}{\partial t} = \Delta[\frac{df(u)}{du} - \theta^2 \Delta u] \quad (1)$$

where $u(x,y,z,t)$ is the concentration of the material and void phases ($-1 \leq u \leq 1$, with $u = -1$ indicating solid material and $u = 1$ indicating void space) at a coordinate *(x,y,z,)*, *t* is the evolution time, $f(u) = \frac{1}{4}(u^2 - 1)^2$ is a double-well free energy function, θ is the width of the interface between the two phases and Δ is the Laplacian operator. The equation was solved in space and time over a cubic domain with edge length of *N*, via a finite difference algorithm. The evolution time controls the characteristic feature size (*λ*), a measure of the domain size of the topology. As time progresses, the features coarsen to reduce interfacial energy, thus increasing *λ*. We choose to extract our topologies at the evolution time corresponding to $\lambda = \frac{1}{5}N$ (loosely corresponding to a sample with 5 × 5 × 5 unit cells), as this number of unit cells was shown in previous studies to provide a homogenized response [46]. To produce spinodal solid topologies with volume fractions different from 50%, a thresholding technique is used. See Appendix A for details.

*2.1.2 Spinodal shell topology*

The spinodal shell topology is extracted from the interface between the solid and void phases, produced as explained in sec. 2.1.1. The volume fraction of reinforcement, $V_f$, is chosen by assigning the appropriate thickness to the shell, $t_s$, so that $V_f = t_s A/V$, with $A$ the surface area of the shell and $V$ the sample volume. The procedure is schematically depicted in Fig. 1 and described in more detail in Appendix A. Spinodal shell topologies extracted from a 50% dense spinodal solid topology have negative Gaussian curvature throughout the shell, with mean curvature close to zero everywhere – thus approaching the geometrical characteristics of TPMSs. While spinodal shell topologies with non-zero mean curvature can be generated by extracting the surface of solid spinodal topologies with densities different from 50%, previous studies have demonstrated that their mechanical properties are inferior [46]. Hence in this work we will limit our attention to spinodal shell topologies with near-zero mean curvature throughout.

*2.1.3 Schwarz P TPMS topology*

The Schwarz P surface was created using the level set approximation technique from the implicit surface:

$$\cos(x) + \cos(y) + \cos(z) = 0 \tag{2}$$

where *(x,y,z)* is a coordinate point in three-dimensional Euclidean space. The minimal surface corresponding to the above equation is generated using *Minisurf*, a software used to generate minimal surface CAD files [53], [54]. Cubic samples with $3 \times 3 \times 3$ unit cells were generated. As explained in sec. 2.1.2, the volume fraction of reinforcement, $V_f$, is chosen by assigning the appropriate thickness to the shell, $t_s$, so that $V_f = t_s A/V$, with $A$ the surface area of the shell and $V$ the sample volume.

*2.1.4 Octet lattice topology*

The octet lattice was built using the commercially available CAD modeling software package SolidWorks. Cubic samples with $3 \times 3 \times 3$ unit cells were generated. The volume fraction of reinforcement, $V_f$, is chosen by assigning the appropriate diameter to the bars, *d*, according to $V_f = 6\sqrt{2}\pi(\frac{d}{l})^2$ with $l$ being the length of the strut [49].

*2.2 Numerical generation of the IPC architectures*

After the reinforcement phase topologies were generated, they were imported into Geomagic Design X, a reverse engineering software that can also be used to handle CAD files. After the appropriate shell thicknesses and bar diameters were generated to result in the desired volume fraction of reinforcement, each CAD model was subtracted from a cube of the same size with a Boolean cut operation to create the complementary matrix phase. Finally, the CAD files are converted into STL files and transferred to the GrabCAD software, the control software for the 3D printer. This process, along with the printed samples, is illustrated in Fig 2.

*2.3 Fabrication approach*

All samples were manufactured with a PolyJet 3D printer (Objet260 Connex3, Stratasys), which allows 3 different materials to be printed simultaneously, with intermediate (digital) formulations obtained by mixing appropriate amounts of the 3 constituent materials at any voxel location. The reinforcement phase in IPCs was printed with VeroWhitePlus, a glassy photopolymer, while the soft matrix phase was printed with Agilus30, a rubbery polymer. All samples are 30x30x30 mm cubes. Selected reinforcement-only cellular materials were printed by replacing Agilus30 with support material, which was subsequently removed by water jetting. All 3D printed samples were left at room temperature for 4 days for curing.

*2.4 Mechanical characterization*

All mechanical tests were performed with an Instron 8800 mechanical test frame equipped with a 100kN load cell. The constituent materials were characterized under both uniaxial tensile and compression loading, using dog bone and cylindrical samples, respectively. Engineering stress and strain were extracted as $\sigma_0 = F/A_0$ and $\varepsilon_0 = \delta/\ell_0$, with $F$ and $\delta$ the raw force and displacement measurements, and $A_0$ and $\ell_0$ the original cross-section and length of the sample, respectively. True stress and true strains were obtained as $\sigma_t = \sigma_0(1 + \varepsilon_0)$ and $\varepsilon_t = \ln(1 + \varepsilon_0)$, respectively. The Young's modulus was extracted as the slope of the initial linear region.

All IPC composites and cellular materials were tested in compression only. ASTM 695-15 for compressive properties of rigid polymer was followed in all compression tests [55]. Engineering stress and strain were extracted as explained above. To minimize the influence of the anisotropic nature of 3D printing introduced by the layer-by-layer printing process, all compression tests were conducted in the direction parallel to the printing direction, which is known to be the strongest. A quasi-static strain rate of 0.001s$^{-1}$ was used in all tests. All samples, except for those subjected to cyclic tests, were compressed to 50% strain to measure Young's modulus, 0.2% offset yield strength, and energy absorption. The latter was extracted as the area under the stress-strain curve up to 50% strain. The cyclic test samples were compressed to 10% strain for three cycles using the same quasi-static strain rate as all in other tests.

*2.5 Finite Elements Modeling*

Finite element meshes of spinodal shell and Schwartz P shell composite samples generated through SimpleWare ScanIP are imported in the commercial Finite Element package Abaqus, while the octet lattice composites are meshed directly within Abaqus. Explicit quasi-static analyses are performed to extract stress-strain curves under uniaxial compression, subject to the following boundary conditions: (i) each composite sample is sandwiched between two frictionless rigid plates; (ii) the top plate moves down and compresses the composite up 25% strain (to ensure quasi-static response, the speed of the top plate is adjusted to ensure that the overall kinetic energy is less 5% of the internal energy throughout the simulation); (iii) all the side faces of the composite sample are left unconstrained.

The reinforcement phase material, VeroWhitePlus, is modeled as an elastoplastic material, with parameters fitted on the experimental results in Fig. 3a. A Young's modulus, $E = 1{,}234$ MPa, is chosen as the average of measured moduli in tension and in compression. The plastic model is chosen to represent the significant tension/compression asymmetry (Fig. 3a): in compression, the response is perfectly plastic, with a yield strength $\sigma_y = 72$ MPa, taken as the average of the plastic flow stress over the plastic strain range of interest; in tension, an initial yield strength of 53 MPa is used, followed by isotropic hardening, $\sigma_y = 53 + 90\ \varepsilon_p^{0.58}$ MPa, with a maximum strength of 62 MPa. Finally, a damage model is used, with a maximum plastic strain damage initiation, $\varepsilon_{po} = 0.017$, and a damage evolution based on linear material softening ($d_p = L \cdot \{\varepsilon_{pf} - \varepsilon_{po}\} = 0.52$, where $d_p$ is the plastic displacement, $L$ is the elemental characteristic length, and $\varepsilon_{pf}$ is the plastic strain at failure).

Since tension/compression asymmetry is small up to ~25% strain, the matrix material, Agilus30, is modeled as a hyperelastic material, with a Marlow strain energy potential fitted to the uniaxial compression engineering stress-strain curve in Fig. 3b.

## 3. Results and Discussion

### *3.1 Mechanical properties of constituent materials*

Constituent materials were first tested in tension and compression as described in sec. 2.4. Stress-strain curves for the two materials are provided in Fig. 3. Notice that the reinforcement phase material is approximately three orders of magnitude stiffer than the matrix phase material; consequently, it is expected that the reinforcement phase will take the vast majority of the load in the composite samples. The matrix phase, which is three orders of magnitude softer than the reinforcement phase, still plays an important role on deformation mechanisms and damage evolution. See Appendix C for details. It is worth noting that both constituent materials show some tension-compression asymmetry, with the reinforcement phase material about 20% stiffer and 32% stronger in compression than in tension. In previous studies, VeroWhitePlus has been shown to exhibit significant size effects on strength [56]. While we observe similar size effects on the tensile properties of VeroWhitePlus dog bone specimen and the compressive properties of VeroWhitePlus cellular samples with spinodal topology (Appendix B), we did not observe any significant size effect in our experimental results on fully dense IPC samples. We attribute this lack of size effects to the fact that all samples experience similar degree of curing regardless of feature size, as the matrix and the reinforcement phases are printed at the same time in fully dense samples, and all samples are of the same size. The conclusion is that size effects on materials properties can be ignored in the following analysis of the mechanical performance of IPC composites. For more details on size effect, refer to Appendix B.

### *3.2 Spinodal IPCs: the difference between solid and shell reinforcement topologies*

We start by comparing the mechanical response of IPCs with spinodal shell and spinodal solid reinforcement topologies, for volume fractions of reinforcement between 20 and 50%. Compressive stress-strain curves are obtained as explained in sec. 2.4, and depicted in Fig. 4a,b. While the response at low relative density is qualitatively similar, IPCs with spinodal shell topologies display a much more gradual failure as the relative density is increased, with much reduced load drops and softening at large strains. Young's modulus ($E$), yield strength ($\sigma_y$) and energy absorption ($U$) are presented as a function of the volume fraction of reinforcement in Fig. 4c,d. For IPCs with spinodal shell reinforcement, we extract the power laws $E \sim V_f^{1.4}$, $\sigma_y \sim V_f^{1.6}$ and $U \sim V_f^{1.1}$, whereas for IPCs with spinodal solid reinforcement we find $E{\sim}V_f^{2.2}$, $\sigma_y{\sim}V_f^{2.7}$ and $U{\sim}V_f^{1.8}$. As the hard reinforcement material is much stronger and stiffer than the rubbery matrix material (with more than three orders of magnitude difference in stiffness, see sec. 3.1), it is expected that the reinforcing phase will dominate the mechanical response of the composite; therefore, the scaling laws obtained here can be compared to those of cellular materials, with the volume fraction of reinforcement representing the relative density, $\bar{\rho}$. For truss-based lattice materials, a scaling $E{\sim}\bar{\rho}^1$ and $\sigma_y{\sim}\bar{\rho}^1$ denote mechanically efficient stretching-dominated behavior, with $E{\sim}\bar{\rho}^2$ and $\sigma_y{\sim}\bar{\rho}^{1.5}$ indicating less efficient bending-dominated behavior [57], [58]. For spinodal cellular materials, $E{\sim}\bar{\rho}^{1.2}$ and $\sigma_y{\sim}\bar{\rho}^{1.2}$ for shell topologies and $E{\sim}\bar{\rho}^{2.3}$ and $\sigma_y{\sim}\bar{\rho}^2$ for solid topologies [46], in good agreement with the scalings found herein for composites, confirming that the reinforcement material dominates the mechanical response. The implication is that IPCs with spinodal shell reinforcement topologies are consistently superior in all metrics, with the advantage increasing significantly at lower volume fractions of reinforcement, where the thin shells with nearly zero mean curvature and negative Gaussian curvatures behave in a predominantly stretching dominated manner and deform uniformly, with little stress intensification. As $V_f$ is increased, the topological difference between solid and shell spinodal reinforcement blurs, and the properties of the two IPCs converge. Given the consistent superior mechanical response of IPCs with spinodal shell reinforcement topology, IPCs with spinodal solid reinforcement topologies are not investigated further.

*3.3 The mechanical advantage of spinodal shell IPC compared to IPCs with regular reinforcement topologies*

It is instructive to compare the mechanical response of IPCs with spinodal shell reinforcement topology, which is intrinsically stochastic, with that of IPCs with periodic reinforcement topologies, in particular the octet lattice topology and the Schwartz P shell topology (Fig. 2). Compressive stress-strain curves obtained over a wide range of volume fractions of reinforcement ($V_f = 5 - 50\%$) are presented in Fig. 5a,b,c. Young's modulus (*E*), yield strength ($\sigma_y$) and energy absorption (*U*) are presented as a function of $V_f$ in Fig. 5d,e,f. While repeat tests at any single density were not conducted, the clear power-law behavior for all properties over the entire density range, with all data points narrowly banded around the average trend, confirms repeatability of the results (e.g., in the case of strength, all data fit within a 98% confidence interval). Two key results clearly emerge: (i) The mechanical properties of IPCs with these three very different reinforcement topologies (periodic truss and shell and stochastic shell) are nearly identical over the entire range of $V_f$, with the spinodal shell IPC performing slightly worse than the others in stiffness and strength (we attribute this to manufacturing defects, as the reinforcement shell of spinodal IPCs is thinner than that of the other two geometries at most volume fractions and is close to the resolution of the 3D printer at 5% volume fraction), and slightly better in energy absorption at high $V_f$; (ii) At $V_f > 35\%$, while the plastic and failure response of octet lattice and Schwartz P shell-based IPCs is qualitatively similar, and characterized by a sharp stress drop immediately after the ultimate strength, the stochastic spinodal shell-based IPC displays a much more gradual failure mechanism, characterized by a nearly flat stress plateau over the entire strain range.

The first result clearly reveals that the bicontinuous nature of the phases is more important than the specific reinforcement topology in determining stiffness, strength and failure initiation of the IPC composite. As many applications require materials to remain within the elastic regime, the implication is that the design space for stiff and strong IPCs is very broad: a wide range of topologies will result in very similar mechanical behavior. Conversely, the post-yielding deformation and failure behavior of the IPC composite (particularly at high volume fraction of reinforcement) is strongly affected by the topological arrangement of the reinforcement and matrix phases, suggesting that topology optimization would play a substantial role in design of IPCs for energy absorption and impact protection.

To better understand this deformation and failure behavior, we compare both the stress-strain curve and deformation response of the IPCs with the three reinforcement topologies, at $V_f = 50\%$ (Fig. 6). Notice that the two types of IPCs with periodic reinforcements experience more catastrophic failure events than the spinodal shell IPC. Soon after the ultimate strength (at $\varepsilon \sim 0.1$), cracks in the reinforcement are clearly visible in the Schwartz P shell IPC; at a strain as low as $\varepsilon \sim 0.2$, these cracks have multiplied and aligned along a shear band, inducing catastrophic failure soon after. While cracks are not visible at the surface of the octet lattice IPC at strains as low as $\varepsilon \sim 0.1$, presumably those cracks exist at the interior of the sample; at $\varepsilon \sim 0.2$, alignment of cracks along a shear band is clearly visible, inducing the same catastrophic failure mechanism as in the Schwartz P shell IPC. In fact, the stress-strain curves of these two IPCs are essentially identical, throughout the entire strain range. By contrast, the stochastic spinodal shell IPC does not exhibit any visible reinforcement cracking until $\varepsilon \sim 0.3$; even then, the cracks appear stochastically distributed across the microstructure, and not banded as for the two periodic topologies. This substantial difference in the failure mechanism results in a nearly flat stress-strain curve up to $\varepsilon \sim 0.3$, followed by a very gentle stress drop up to $\varepsilon \sim 0.4$, when the test was interrupted.

To further investigate the damage resistance of the spinodal shell IPC, loading-unloading compression experiments were performed, whereby IPCs with the three reinforcement topologies, at $V_f = 50$, were compressed to 10% strain for three cycles (Fig. 7). As the number of cycles increases, all IPCs show increasingly visible cracks in the reinforcement phase. The Schwartz P shell IPC and the octet lattice IPC both experience decreasing load-bearing capacity as a result. On the contrary, even with increasing visible fracture sites, the load-bearing capacity of the spinodal shell IPC is largely unaffected by the cyclic loading, demonstrating superior resistance to damage.

We attribute these dramatic differences in deformation and failure behavior to (i) the larger surface area and (ii) the stochastic nature of the spinodal shell topology. The larger surface area of the spinodal shell compared to the other two periodic geometries results in increased reinforcement/matrix interaction, with the matrix preventing/arresting crack propagation in the reinforcement phase. For a more detailed discussion on the influence of surface area, refer to Appendix D. In addition, the stochastic nature and complex shape of the spinodal shell reinforcement distribute the loads more efficiently throughout the topology, maintaining load-bearing capacity after fracture initiation and preventing the formation of catastrophic failure-inducing crack bands. In periodic geometries like Schwartz P or octet truss, failures first occur around stress concentration locations, and subsequently band along specific directions, leading to catastrophic failure and loss of load bearing capacity. Conversely, for the stochastic spinodal shell topology, stress is more uniformly distributed across the entire structure. Even after fracture of some shell sections, the complex topology still allows the spinodal shell-reinforced composite to maintain nearly unchanged load bearing capacity.

The difference in failure response between the spinodal and the two types of periodic reinforcement IPCs is less pronounced at low volume fractions of the reinforcement (Fig. 5), as the increased volume fraction of the matrix helps stabilize even periodic structures against crack banding of the reinforcement.

To better understand and quantify the differences in deformation and failure behavior, finite element analyses were performed on the IPCs with the three reinforcement topologies, at a volume fraction of reinforcement $V_f = 30\%$. The results are presented in Fig. 8. Notice that the computational prediction of the stress-strain response is in good agreement with the experimental results, well capturing the initial stiffness, the yield and ultimate strength and the beginning of the post-failure behavior (Fig. 8a,b,c). This agreement validates the computational model and provide confidence in its ability to capture the onset and early evolution of damage. As finite elements simulations involving post-failure behavior (including fracture) are very challenging because of stress singularity and loss of uniqueness, a full numerical description of the failure mechanisms all the way to densification is beyond the scope of this work. Nonetheless, several conclusions can be extracted from an analysis of the stress state at the early and intermediate phases of deformation. First we compare the von Mises stress distribution in the reinforcement phase for all three IPCs at a strain of 0.1, roughly coinciding with the attainment of the maximum strength, in Fig. 8d,e,f. It is apparent that the spinodal shell IPC shows small and uniformly distributed stress concentrations throughout the sample, whereas the two IPCs with the periodic reinforcement topologies exhibit very large and interconnected stress concentrations. This is consistent with the experimental findings in Fig. 6 and further supports the argument that the spinodal shell IPC is effective at avoiding catastrophic failure by preventing formation of reinforcement cracking bands. On the contrary, in IPCs with periodic reinforcement such as the octet and Schwartz P IPC studied here, fracture of a member leads to a cascading crack propagation which results in a near-instant loss of load bearing capacity. We then compare the von Mises stress distribution in the reinforcement phase in the middle of the sample for all three IPCs at a strain of 0.15, when surface cracks start to appear in Fig. 8g,h,I. It can be seen that the stress distribution of the spinodal shell IPC is still uniform and largely unaffected by cracks, while the two periodic IPCs experience drop in load bearing capacity as a result of cracks at stress concentration points. These simulation results are also in good agreement with the cyclic experiments shown in Fig. 7 and demonstrate the excellent damage resistance of the spinodal shell IPC. Movies of the evolution of the von Mises stress distribution in the reinforcement phase for all three IPCs generated in the simulation can be found in Appendix E: Supplementary Data.

Collectively, these experimental and numerical results clearly illustrate that the stochastic nature of spinodal shell topologies enables the establishment of a uniform stress field throughout the reinforcement, which persists even after the onset of reinforcement cracking. The lack of substantial stress intensifications promotes a stochastic distribution of initial cracking locations and prevents the catastrophic occurrence of crack banding, resulting in a flat stress plateau through very large strains. Combined with a very large interfacial area, which ensures intimate reinforcement/matrix interaction, this feature makes spinodal shell IPCs ideally suited for energy absorption and impact applications.

## 4. Conclusions

In summary, we have fabricated and mechanically investigated a new type of interpenetrating phase composite (IPC) with spinodal shell reinforcement topology, comparing their mechanical response to that of well-established mechanically efficient periodic IPCs with octet lattice and Schwartz P shell reinforcement topologies. Polymeric model systems were produced by Polyjet multi-material additive manufacturing; while the mechanical properties of the two polymeric phases (and hence the resulting composites) were inferior to those of any practical structural material, this technique allowed unbiased comparison of different topologies. We have shown that while all three types of IPCs perform nearly identically in terms of initial stiffness, yield strength and energy absorption over a wide range of volume fraction of reinforcement (5-50%), spinodal shell IPCs are far more robust than any other IPCs, exhibiting greater damage resistance as well as a much more uniform deformation and gradual failure. This unique feature is attributed to: (1) uniform distribution of stresses and strains in shell topologies, stemming from fairly uniform distribution of negative Gaussian curvature across the entire surface; (2) larger surface area at a given volume fraction of reinforcement, resulting in increased matrix support on the load bearing reinforcing phase; (3) the stochastic nature and complex shape of spinodal topologies, which not only act as crack barriers and locally inhibit crack propagation and banding, but also continue to provide load-bearing capacity even after fractures of some members occurs. The combination of excellent mechanical efficiency, on par with those of the best IPCs with periodic reinforcement, great damage resistance, gradual deformation and failure mechanisms and potential for scalable manufacturing makes spinodal shell IPCs exceptional candidates for damage tolerance and energy absorption applications where a prolonged compressive stress plateau after maximum stress is desired. While quantitative assessment of impact performance requires high-strain rate testing and are beyond the scope of this work, previous studies have demonstrated that impact performance ranking of cellular materials can be obtained by the quasi-static experiments performed in this work [59], [60]. This strongly suggests that spinodal shell IPCs would be excellent performers for impact protection.

Most importantly, unlike periodic IPCs, spinodal shell IPCs can in principle be scalably manufactured at various length scales, using self-assembly approaches followed by material conversion techniques. Possible self-assembly approaches include spinodal decomposition of block copolymers [45], interfacially jammed colloidal suspensions (bijels) [40], [44] and selective etching of bimetallic alloys [41]. These approaches allow ready fabrication of polymeric, metallic or ceramic macro-scale samples with domain sizes at the micro or nano-scale, resulting in architected materials with enormous surface area and further improving mechanical properties by virtue of well-established size effects on the constituent materials [61]–[63]. As spinodal topologies are bicontinuous, such micro/nano-architected materials could be infiltrated with a second phase via deposition or infiltration processes, potentially providing a uniquely scalable fabrication process for shell-based IPCs.

Finally, we emphasize that, while the multi-material additive manufacturing approach used in this study resulted in polymer-polymer composites with absolute mechanical properties far inferior to those of any structural material, the self-assembly-based processes envisioned above could be used to produce ceramic-polymer, ceramic-metal, metal-polymer and metal-metal composites that can in principle outperform most existing structural materials. Demonstration and characterization of such advanced spinodal shell-based IPCs will be the subject of future studies.

**Appendix A. Generation of spinodal topologies**

The generation of spinodal topologies follows an approach discussed in detail in [42,51], and summarized here for completeness. Spinodal decomposition can be described by the Cahn-Hillard evolution equation [37]:

$$\frac{\partial u}{\partial t} = \Delta[\frac{df(u)}{du} - \theta^2 \Delta u] \qquad (A1)$$

where $u(x,y,z,t)$ is the concentration of the two phases A and B at a coordinate *(x,y,z)* ($-1 \leq u \leq 1$, with $u = -1$ indicating only phase A, or void space, and $u = 1$ indicating only phase B, or solid material), *t* is the decomposition time, $f(u) = \frac{1}{4}(u^2 - 1)^2$ is a double-well free energy function, θ is the width of the interface between the two phases and $\Delta$ is the Laplacian operator. Equation (A1) is solved numerically with a finite difference scheme over a cubic volume with edge length N = 100, which is discretized into a lattice of mesh size, ℓ=N/100=1. Let $u_{ijk}^m$ denote the discrete value of the phase field variable $u(i,j,k,m\tau)$ at nodal point *(i,j,k)*, with $\tau$ the integration time step, chosen to be sufficiently small to achieve convergence ($\tau$ =0.005 was used here), and *m* the time step. After discretization with a finite difference scheme, equation (A1) can be written as:

$$\frac{u_{ijk}^{m+1} - u_{ijk}^m}{\tau} = \Delta[(u_{ijk}^m)^3 - u_{ijk}^m - \theta^2 \Delta u_{ijk}^m] \qquad (A2)$$

where $\theta$ is the thickness of the interface between the two phases and $\Delta$ is the Laplacian operator. The following boundary conditions are applied to solve equation (A2):

$u(i,j,k,m\tau)=u(i+L,j,k,m\tau)$ (A3.1)

$u(i,j,k,m\tau)=u(i,j+L,k,m\tau)$ (A3.2)

$$u(i,j,k,m\tau)=u(i,j,k+L,m\tau) \tag{A3.3}$$

A randomly generated initial condition, $u(i,j,k,0) = u_0(i,j,k,0) \in [-5,5] \times 10^{-4} \neq 0$, was used as a perturbation to exit unstable equilibrium and start the decomposition kinetics. As the solution progresses, the system phase separates at early times, and subsequently continues to coarsen; during the coarsening phase, the curvature of the interface between solid and void decreases and the size of the single-phase domains increases. A cutoff $u_c^m$ is defined to separate phase A from phase B, with the phase at a point $(i,j,k)$ and a time $t = m\tau$ defined as:

$$G_{ijk}^m = H(u_{ijk}^m - u_c^m) \tag{A4}$$

where $H(\cdot)$ represents the Heaviside function. To achieve a 50% volume fraction ($V$) of phase A, the cutoff $u_c^m$ is adjusted so that $u_{ijk}^m$ satisfied the distribution given by:

$$G_c^m = \frac{1}{N^3}\Sigma_{i=1}^N \Sigma_{j=1}^N \Sigma_{k=1}^N G_{ijk}^m = \frac{1}{N^3}\Sigma_{i=1}^N \Sigma_{j=1}^N \Sigma_{k=1}^N H(u_{ijk}^m - u_c^m) = V \tag{A5}$$

Here the decomposition time that controls the characteristic feature size ($\lambda$) has been set to provide $\lambda = 1/5\ N$, a geometrical condition that has been shown to provide the best overall mechanical performance for spinodal shell-based cellular materials [46].

The spinodal shell topologies are subsequently derived by extracting the surface from spinodal solid cellular topologies with volume fraction, $V$=50%, resulting in a shell topology with negative Gaussian curvature and near-zero mean curvature throughout the domain. While shell-topologies with non-zero mean curvature can be derived from spinodal solid topologies with $V \neq 50\%$, previous studies have demonstrated that near-zero mean curvature topologies provide the best overall mechanical performance for spinodal shell-based cellular materials [46].

**Appendix B. Size effect on the mechanical properties of 3D printed consituent materials**

The mechanical properties of the photopolymers used in this study, namely the glassy VeroWhitePlus and the rubbery Agilus30, have not been as thoroughly investigated as other commonly used structural materials for additive manufacturing. As several materials for additive manufacturing are known to exhibit size-dependent mechanical properties, we investigated the tensile properties of dog bone samples printed with VeroWhitePlus (the reinforcement material in our study), as a function of sample size. The results are reported in Fig. B.1a, and show a clear size effect for sample dimensions lower than ~1mm, with smaller samples being less stiff and strong, but far more ductile. The results are consistent with a recent study, which has shown a well-defined size threshold for the strength of VeroWhitePlus: above this threshold, no size effect is observed, whereas below this threshold, a smaller-is-weaker effect is present [56]. This size threshold was estimated to be approximately 50 times the printing layer thickness, which is in good agreement with the threshold identified in our dog bone experiments. The size threshold is shown as a horizontal line in Fig. B.2, overlaid with the reinforcement feature sizes of IPC samples tested in this study. According to Fig. B.2, we would expect to see no size effect for most IPCs examined in this study, with the exception of IPCs with spinodal shell topologies for reinforcement volume fractions smaller than 30%. To verify this assumption, we proceeded to test in compression two 30% dense spinodal shell cellular materials, with identical topology but different shell thicknesses, $t_1 = 0.75mm$ and $t_2 = 1.5mm$. The latter was printed with two times the overall dimension of the former, so to have the same topology and volume fraction but different shell thickness. As shown in Fig. B.1b, a significant size effect exists, quantitatively consistent with the dog bone results in Fig. B.1a and the threshold depicted in Fig. B.2. Finally, we tested in compression two spinodal shell IPCs, identical to those in Fig. B.1b but with the void space occupied by the Agilus30 matrix. The results, reported in Fig. B.1c, remarkably show effectively no size effect on stiffness or strength, albeit the effect on ductility is pronounced. This lack of size effect is also visible in Fig. 5 in the main manuscript, where no change of slope in the strength VS volume fraction curves is observed for spinodal shell composites, in disagreement with the predictions of Fig. B.2. This apparent disagreement can be explained as follows. A recent study on size effect in 3D printing polymers has shown that the size effect observed on thinner samples emanate from the layer-by-layer printing process [64]. As the sample thickness increases, previously printed layers are exposed longer to UV light, which leads to higher curing degree and better mechanical properties. In agreement with this study, we believe that the absence of size effect in our IPC composites stems from the fact that the reinforcing phase and the matrix phase are printed on the same layer in the same scan over the entire sample, so that printing composites is qualitatively similar to printing solid blocks. As all composite samples have similar size and therefore similar number of layers, the polymer in them is cured to a similar degree.

The important implication is that size effects on the base material properties, while generally important, can be ignored in the analysis of the mechanical properties of the IPC composites discussed in this study.

**Appendix C. Comparison of the deformation mechanisms in composite and cellular (reinforcement only) samples**

Past research has shown that the soft phase in a 3D printed composite helps stabilize the reinforcing cellular materials which leads to more uniform deformation and increased load bearing capacity [34]. Fig. C.1 illustrates the deformation mechanisms of IPC and cellular (i.e., reinforcement-only) samples, for all three reinforcement topologies, i.e. octet lattice, Schwartz P and spinodal shell. For all three reinforcement topologies, the general finding is that reinforcement-only samples fail in a more localized and catastrophic fashion, whereas the deformation of IPC samples is stabilized by the matrix, which distributes stress/strain more uniformly across the samples and arrests crack propagation upon individual member fracture, ultimately delaying failure. Interestingly, while the spinodal shell reinforcement-only sample deforms in a more uniform manner compared to the other two periodic geometries, the addition of the matrix still helps stabilize the reinforcement shell, and changes the failure mode of the spinodal shell IPC from shell bending and buckling (as observed in the reinforcement-only sample, Fig. C.1e) to shell fracture (Fig. C.1f). This matrix contribution is attributed to the large surface area of the spinodal shell topology and the incompressibility of the matrix: while larger surface area leads to increased support from the matrix, incompressibility promotes membrane deformation of the shell members, while preventing bending and buckling. The overarching conclusion is that the matrix, which is three orders of magnitude softer than the reinforcement, barely contributes to initial stiffness and strength of the composite, but has a pronounced effect on deformation mechanisms and damage evolution.

**Appendix D. Influence of surface area of the reinforcement phase in IPCs**

One possible reason for the unique properties of spinodal shell IPCs is the increased surface area of the spinodal shell relative to the other reinforcement topologies, which leads to increased interaction and support from the matrix. As discussed in Appendix A, the surface area and curvature of spinodal geometry is affected by decomposition time $t$, with increasing $t$ resulting in deceased surface area and curvature. The decomposition time $t$ can be controlled to generate different ratio between the characteristic length over the cubic length, $\lambda/L$ (Fig. D. 2). In all spinodal samples tested in the manuscript, $\lambda/L = 1/5$ was used as it has been shown to provide the best overall mechanical properties to spinodal shell cellular materials[46]. The surface area of the 3D printing models for all three topologies was extracted in Netfabb, a 3D printing software, and plotted in Fig. D.1. Notice that the spinodal shell topology has significantly more surface area than both Schwartz P and octet lattice topologies, especially at lower volume fractions of reinforcement. As seen in Fig. 6, even at high volume fraction, where the octet lattice has similar surface area, spinodal shell IPCs are still better at avoiding catastrophic failures and have better energy absorption as a result. This is attributed to the stochastic nature and complex shape of the spinodal shell topology.

The influence of curvature and surface area on mechanical properties of spinodal shell IPCs deserves further investigation. Some preliminary results are shown in Fig. D. 3. Samples with $\lambda/L = 1/8$ have the highest surface area and curvature, while samples with $\lambda/L = 1/3$ have the lowest. It can be seen that samples with $\lambda/L = 1/8$ seem to perform slightly better mechanically than samples with $\lambda/L = 1/5$, and significantly better than samples $\lambda/L = 1/3$. One possible explanation is that samples with $\lambda/L = 1/3$ are more susceptible to defects from manufacturing. It has been found in our previous study that samples with $\lambda/L = 1/8$ are less sensitive to imperfection than samples with $\lambda/L = 1/3$ because of its increased surface area and more stochastic nature [46]. Further experiments and simulation are needed to better understand this phenomenon.

**Appendix E. Supplementary Data**

Movie S1. Numerical simulation of deformation and fracture of spinodal shell-reinforced IPC, under 10% compressive strain. Contours represent the Von Mises stress distribution.

Movie S2. Numerical simulation of deformation and fracture of Schwartz P shell-reinforced IPC, under 10% compressive strain. Contours represent the Von Mises stress distribution.

Movie S3. Numerical simulation of deformation and fracture of octet lattice-reinforced IPC, under 10% compressive strain. Contours represent the Von Mises stress distribution.

**Acknowledgments**

This work was supported by the Office of Naval Research (program Manager: D. Shifler, Grant No. N00014-17-1-2874). The ABAQUS Finite Element Analysis software is licensed from Dassault Systemes SIMULIA, as part of a

# Figures

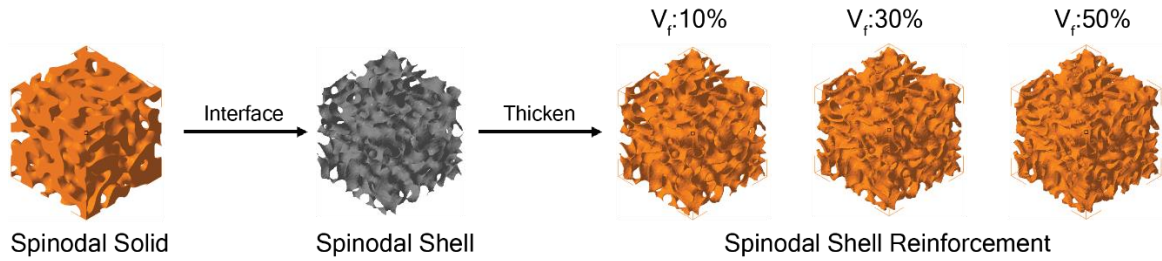

**Fig 1.** Schematic of approach for numerically generating the spinodal shell reinforcement phase for IPCs.

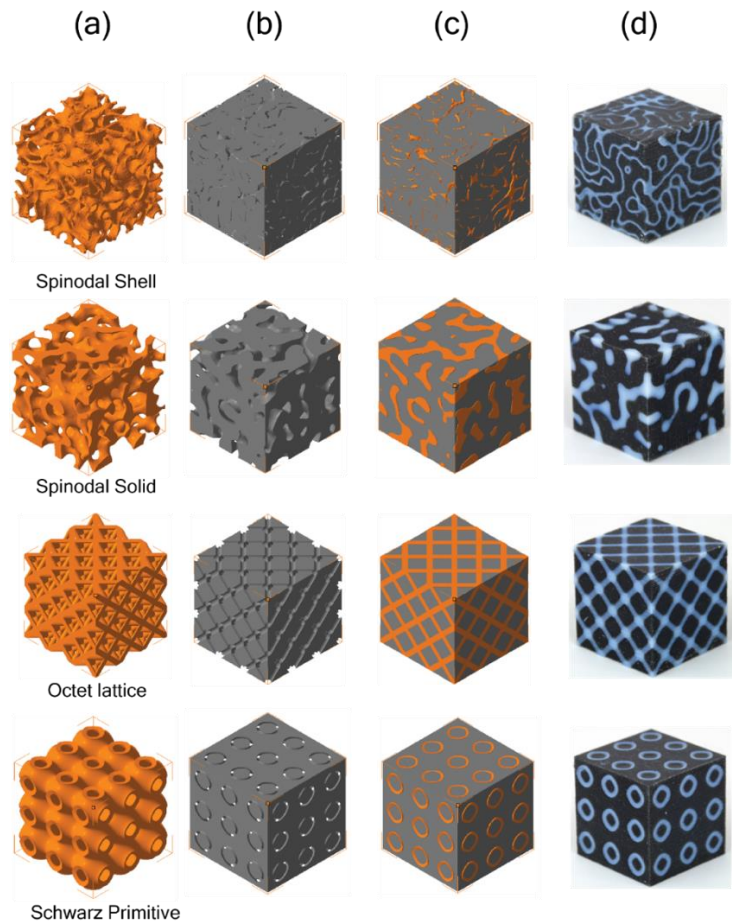

**Fig 2.** (a) Reinforcement phase, (b) matrix phase, (c) assembled interpenetrating phase compositesand (d) 3D printed composite sample for spinodal shell, spinodal solid, octet lattice and Schwarz P shell IPCs.

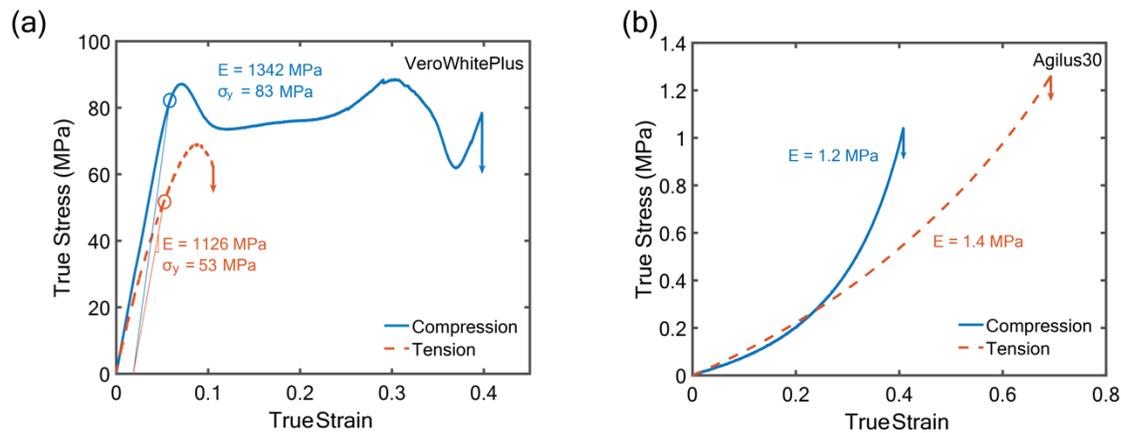

**Fig 3.** Tensile and compressive true stress – true strain curves for (a) VeroWhitePlus and (b) Agilus30.

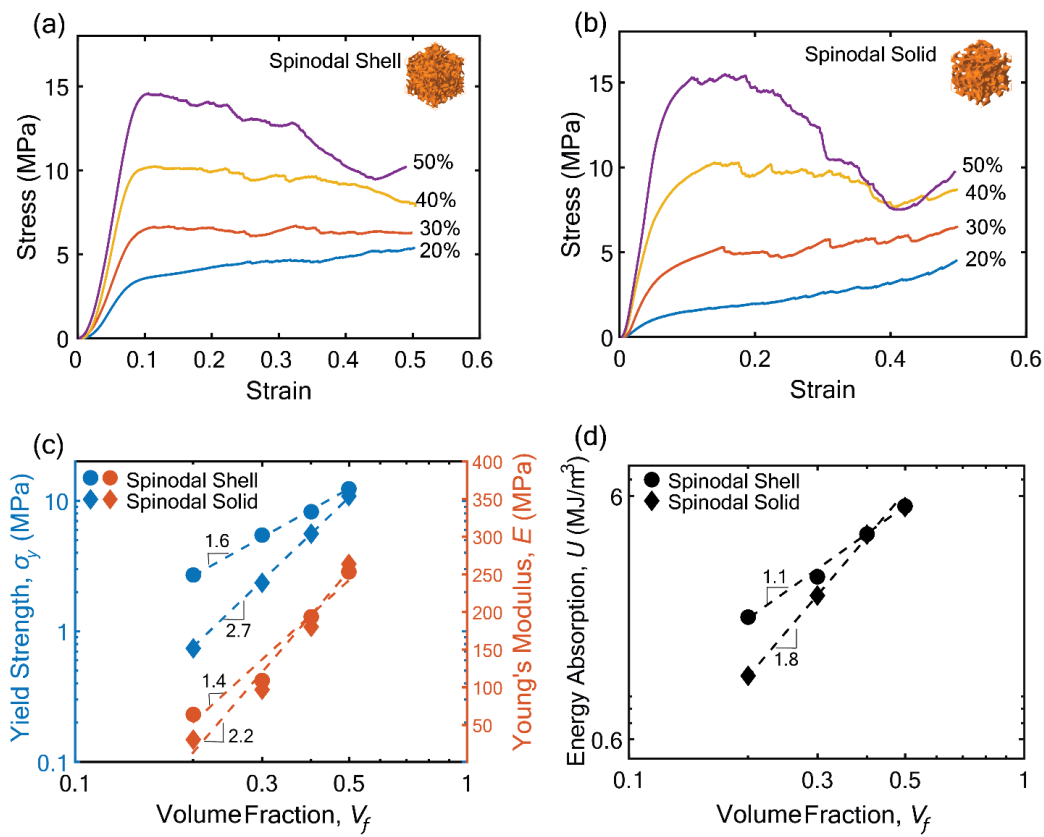

**Fig 4.** Comparison between the mechanical response of IPCs with spinodal shell reinforcement and spinodal solid reinforcement. (a, b) Compressive stress-strain curves. The volume fraction of reinforcement ranges from 20% to 50%. (a) IPCs with spinodal shell reinforcement. (b) IPCs with spinodal solid reinforcement. (c, d) Mechanical properties as a function of volume fraction of reinforcement phase: (c) Young's modulus and yield strength; (d) Energy absorption.

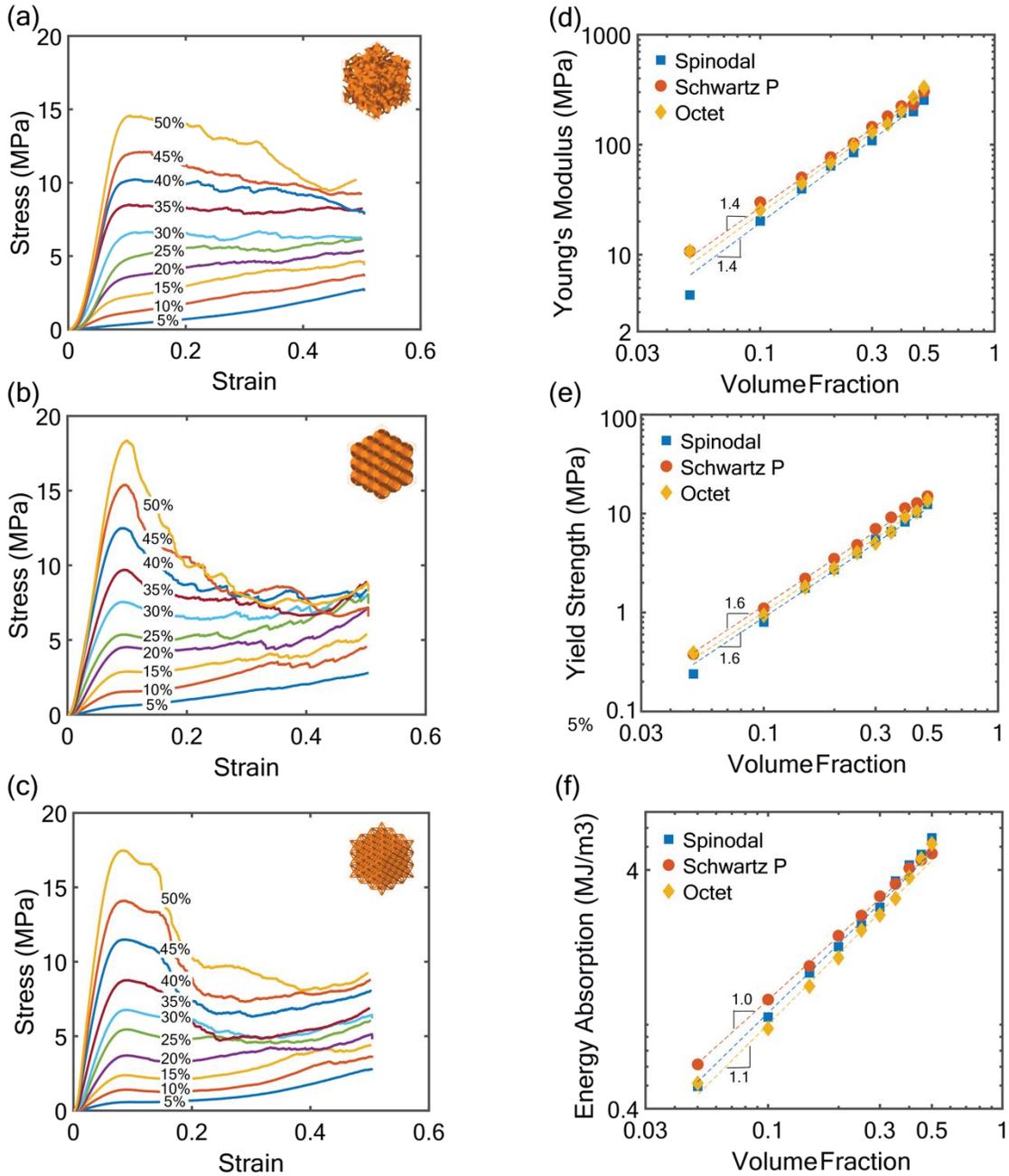

**Fig 5.** (a-c) Stress-strain curves of IPCs with different reinforcement topologies at various volume fractions compressed to 50% strain. Notice that spinodal shell IPCs don't exhibit catastrophic load drops at high volume fractions of reinforcement. (d- f) Comparison of mechanical properties of IPCs at different reinforcement volume fractions: (d) Young's modulus; (e) Yield strength; (f) Energy absorption. Mechanical properties of spinodal IPC at 5% were excluded from the scaling as the thickness of the reinforcement shell is close to the resolution of the 3D printer and is susceptible to manufacturing defects.

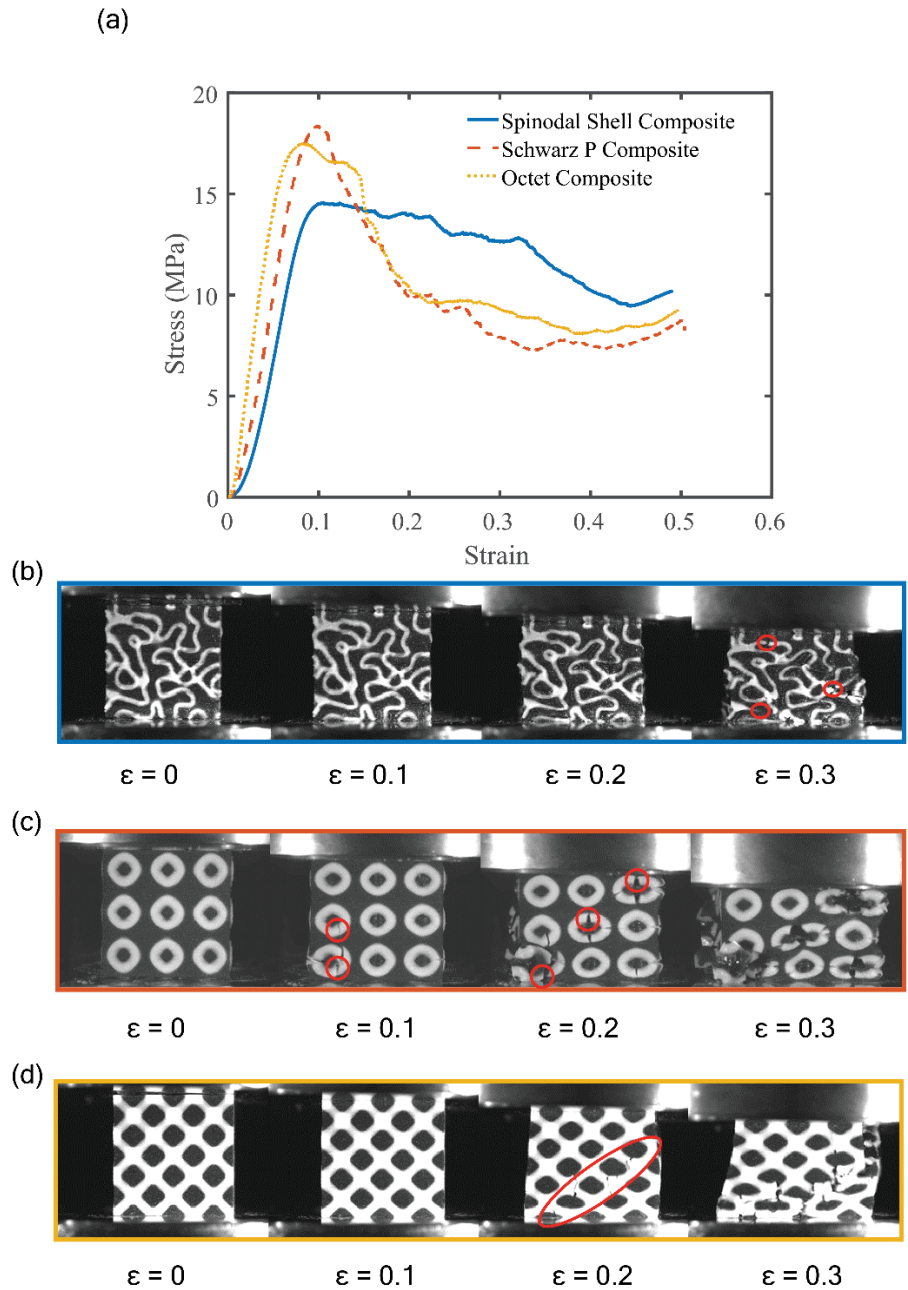

**Fig 6.** Deformation response of IPCs with different reinforcement topologies at 50% volume fractions, compressed to 50% strain. (a) Stress-strain curve. (b) Spinodal shell composites. (c) Schwartz P shell composites. (d) Octet lattice composites. Notice that octet and Schwartz P IPCs at high volume fraction of reinforcement experience more localized catastrophic failures than spinodal shell IPCs.

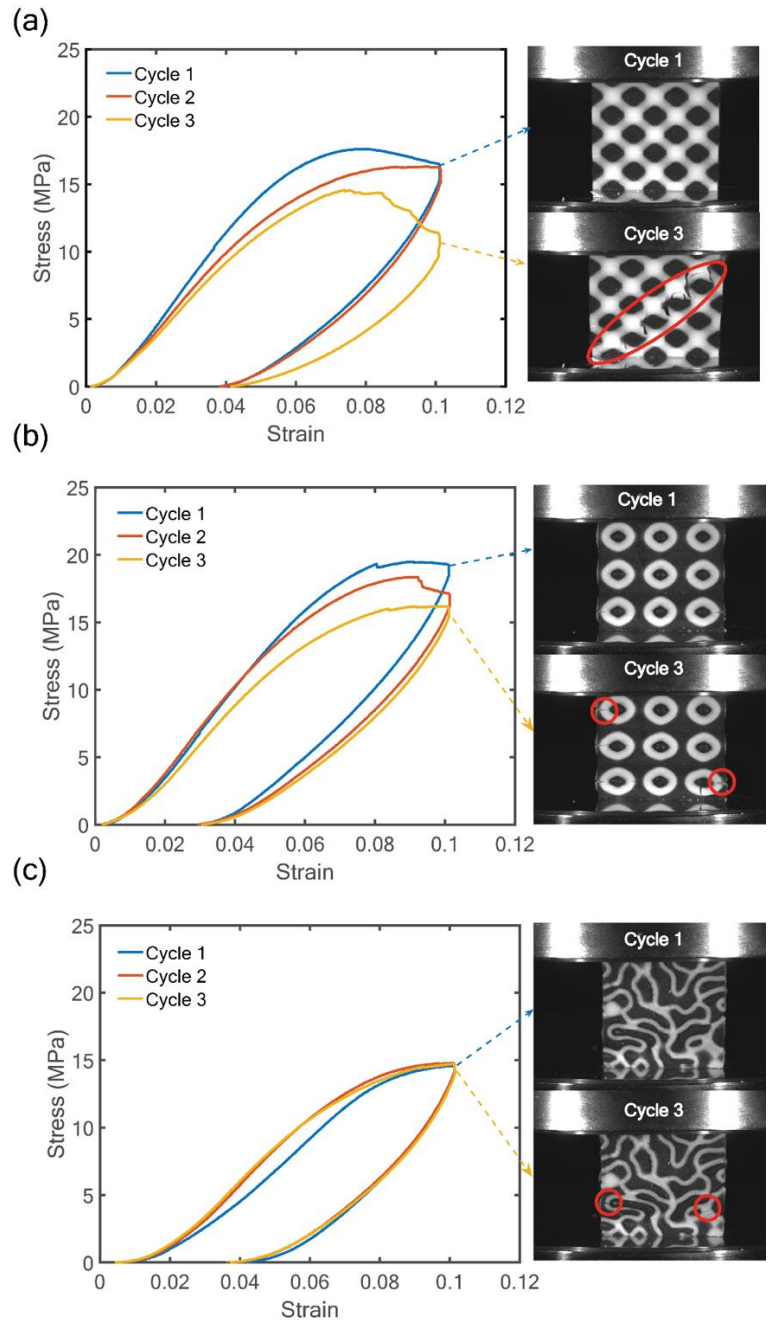

**Fig 7.** Cyclic compression experiments on IPCs with different reinforcement topologies, at 50% volume fraction of reinforcement. All samples have been compressed to 10% strain for 3 cycles. (a) Octet lattice composite. (b) Schwartz P shell composite. (c) Spinodal shell composite. All three topologies show increasingly visible fractures in the reinforcement phase as cycling loading progresses. Both periodic IPCs show decreasing load bearing capacity as a result, while the stress-strain curve of the spinodal shell composite is largely unaffected.

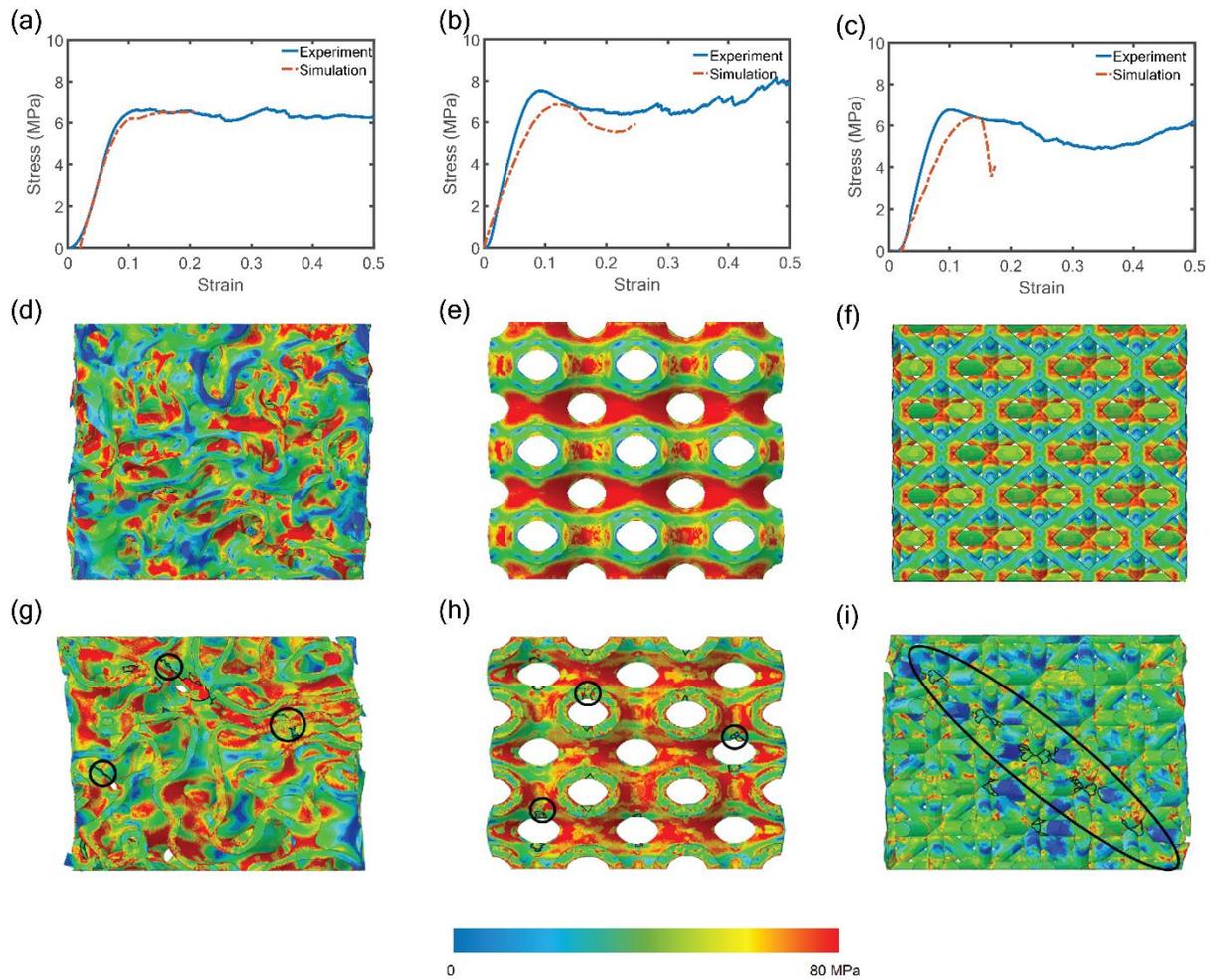

**Fig 8.** Comparison between experiment and simulation results. (a, b, c) Stress strain curve comparison between experiment and simulations, for IPCs at 30% volume fraction of reinforcement: (a) spinodal shell IPC; (b) Schwartz P shell IPC; (c) octet lattice IPC. (d, e, f) von Mises stress map of the reinforcement phase, extracted from the composite at 10% strain: (d) spinodal shell reinforcement; (e) Schwartz P shell reinforcement; (f) octet lattice reinforcement. (g, h, i) von Mises stress map of the reinforcement phase at the half-way cross section, extracted from the composite at 15% strain: (g) spinodal shell reinforcement; (h) Schwartz P shell reinforcement; (i) octet lattice reinforcement. Cracks are highlighted in black.

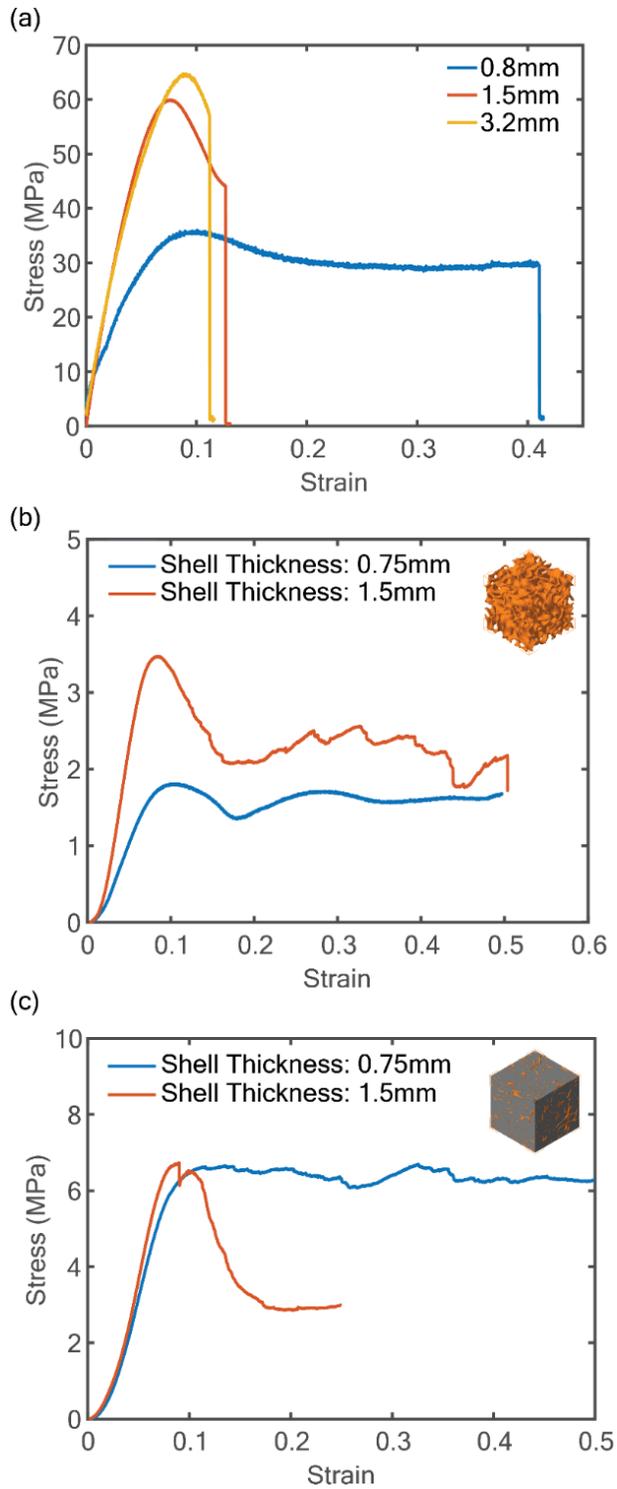

**Fig. B.1.** (a) Stress-strain curves of tensile samples with different thickness; (b) Compressive stress-strain curves of spinodal shell cellular (reinforcement only) sample at 30% volume fraction of reinforcement, for different shell thicknesses; (c) Compressive stress-strain curves of spinodal shell IPC at 30% volume fraction of reinforcement, for different shell thicknesses.

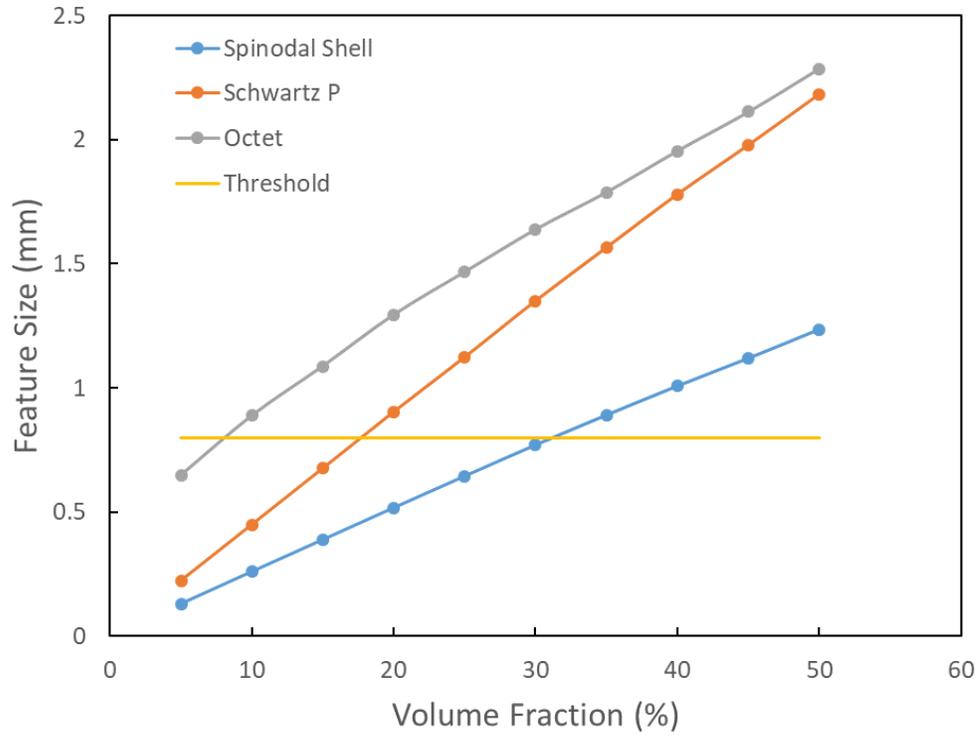

**Fig. B.2.** Feature size of the reinforcing phase as a function of the volume fraction of reinforcement, for different topologies. The yellow horizontal line represents the threshold below which size effects are expected.

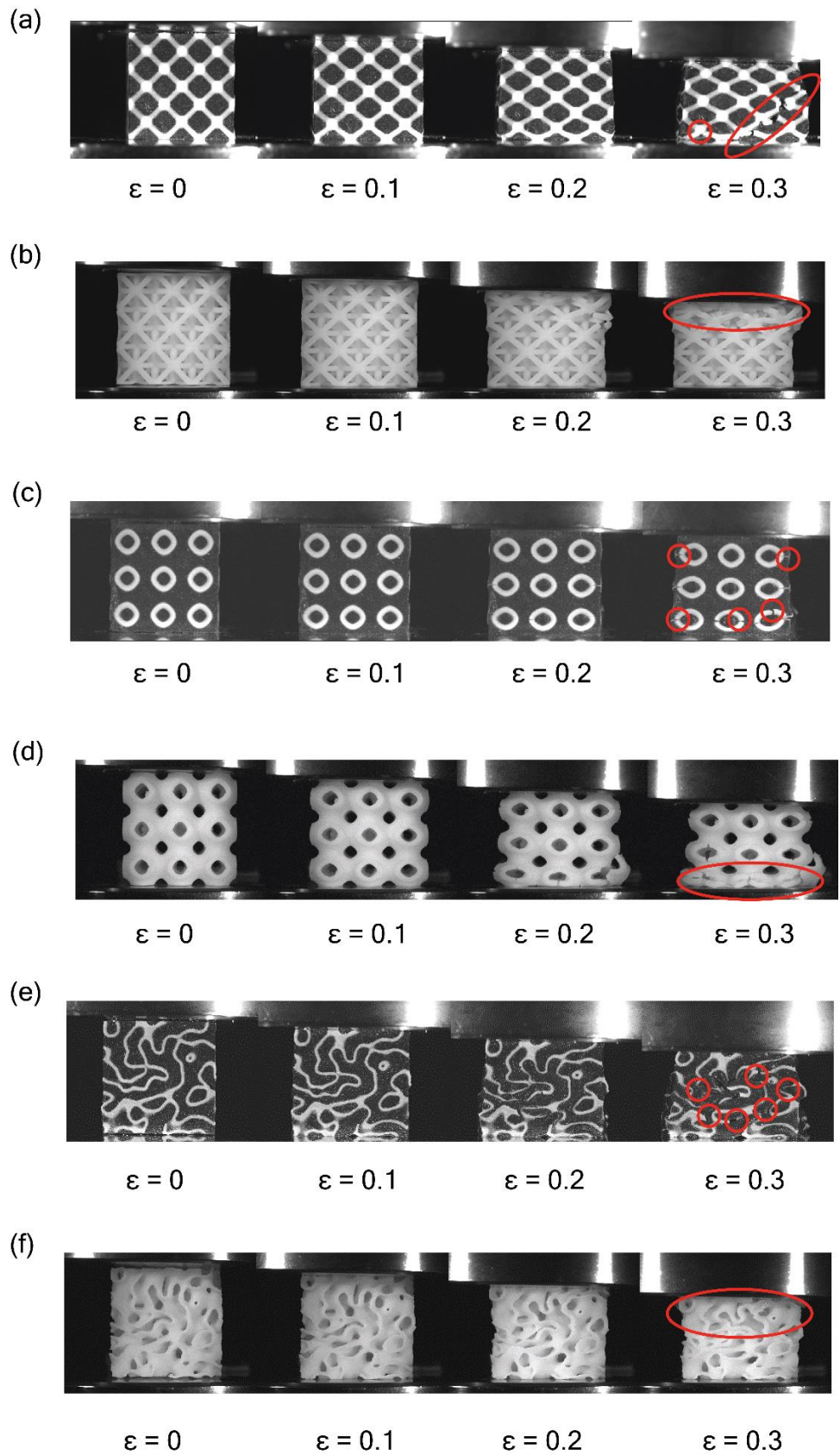

**Fig. C.1.** Deformation mechanisms comparison between composite and cellular (reinforcement only) samples with 30% volume fraction at different strain levels: (a) Octet lattice composite, (b) Octet lattice reinforcement, (c) Schwartz P shell composite, (d) Schwartz P shell reinforcement, (e) Spinodal shell composite, (f) Spinodal shell reinforcement. For all reinforcement topologies, the composite shows more uniform and less catastrophic deformation, while the reinforcement only sample shows more localized deformation and failures. Addition of matrix also changed the failure mode of spinodal shell composite from shell bending or buckling of the reinforcement only sample to failure from shell fracture because of the incompressibility and support of the matrix.

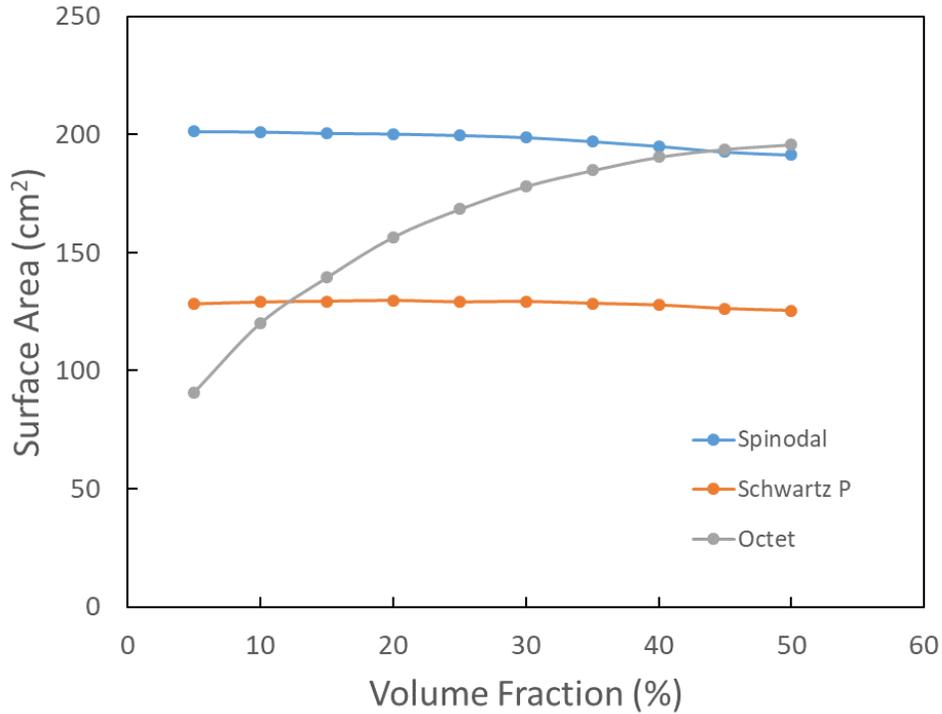

**Fig. D. 1**. Surface area of the reinforcing phase for IPC samples with different reinforcement topologies.

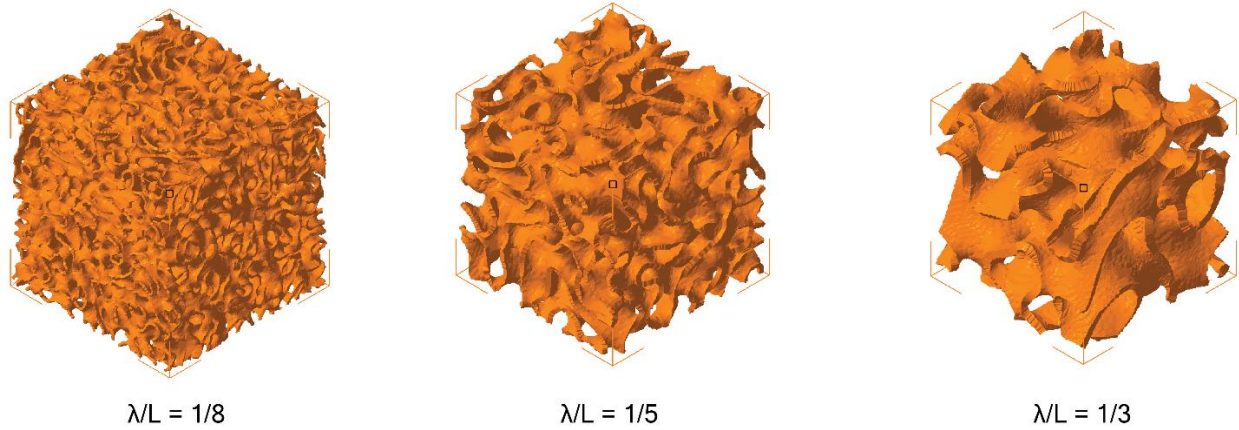

**Fig. D. 2.** $\lambda/L = 1/8$, $\lambda/L = 1/5$ and $\lambda/L = 1/3$ spinodal shell cellular materials at 30% volume fraction

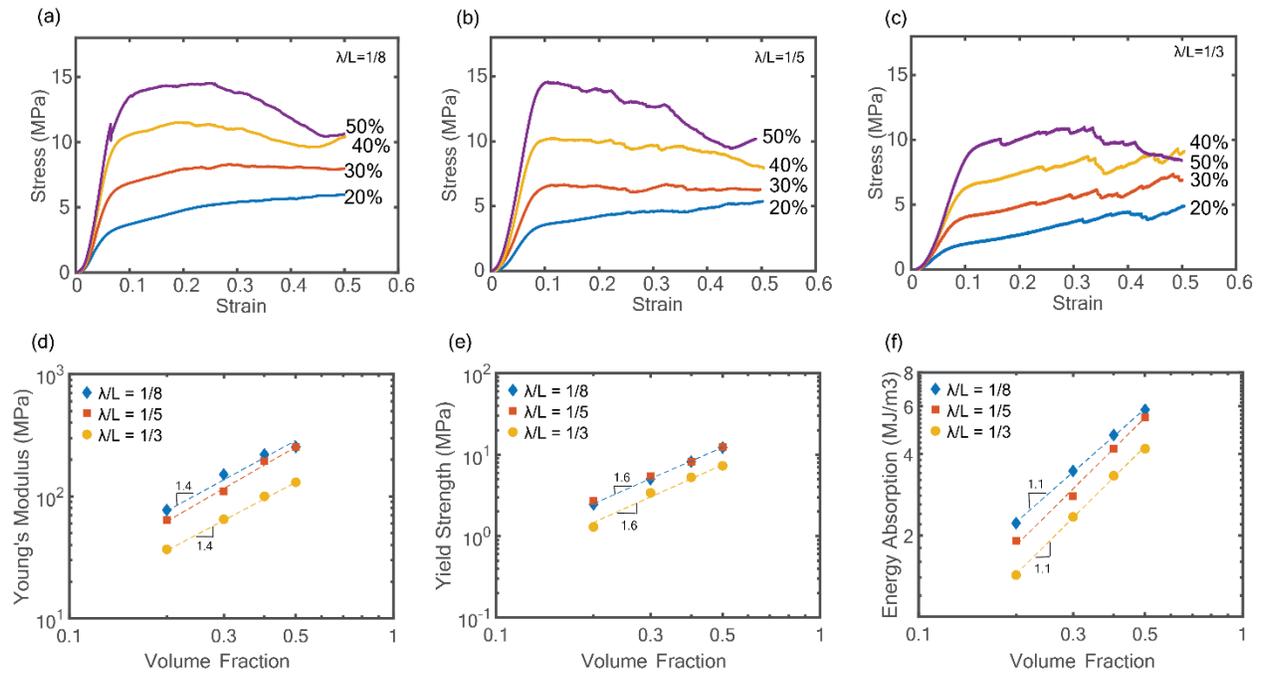

**Fig. D. 3.** Mechanical properties of spinodal shell IPCs with different curvature/surface area. Stress-strain curves for (a) $\lambda/L = 1/8$; (b) $\lambda/L = 1/5$; (c) $\lambda/L = 1/3$. (d) Young's modulus. (e) Yield strength. (f) Energy absorption.